\documentclass[12pt]{article}
\let\section=\subsection     \let\subsection=\subsubsection                
\usepackage{epsfig}
\usepackage{amsmath}
\begin{document}
%
\begin{center}
   {\large \bf QCD Thermodynamics and Fireball}\\[2mm]
   {\large \bf Evolution in URHICs\footnote{Work supported in part by BMBF and GSI.}}\\[5mm]
   T.~Renk$^a$, R.~A.~Schneider$^a$, and W.~Weise$^{ab}$ \\[5mm]
   {\small \it  $^a$ Physik-Department, Technische Universit\"at M\"unchen,\\
   85747 Garching/M\"unchen, Germany \\[2mm]
   $^b$ECT*, Villa Tambosi, 38050 Villazzano (Trento), Italy \\[8mm]}
\end{center}
%
%
\begin{abstract}\noindent
The fireball created in an ultrarelativistic heavy ion collision
is the environment in which all processes providing clues about
the possible formation of the quark-gluon plasma (QGP) happen. It is therefore
crucial to understand the dynamics of this hot and dense system.
We set up a model in which the fireball evolution is reconstructed
between two stages, the freeze-out, which is accessible by hadronic
observables, and the initial collision for which the overlap geometry
can be calculated. Using the equation of state (EoS) provided by
a quasiparticle model of the QGP, we are able to calculate
thermodynamical properties in volume slices of constant proper time
and determine the volume expansion self-consistently.
The resulting evolution model can then be tested against other
observables, such as dilepton yields.
\end{abstract}
%
%
%
\section{Introduction}
%
One of the most striking predictions of QCD at high temperatures is
the formation of the quark-gluon plasma (QGP), a deconfined
phase with quarks and gluons as degrees of freedom. 
Unfortunately, finding direct evidence for this phase
or studying its properties is not an easy task. Ideally, one
would like to study the behaviour of an observable sensitive
to the formation of the QGP, such as the J/$\Psi$ system
or dilepton radiation. In reality, however, no static QGP system can be
prepared in a heavy ion collision and all processes
inside the resulting fireball are necessarily convoluted with its evolution dynamics.
As this leaves two a priori unknown inputs into
any model calculation, the modified reaction
dynamics inside the system and the fireball evolution itself, 
a model which is able to
describe just one (or a few) observables may not be sufficient
to disentangle these two pieces and extract unambiguous
information.

The proper way to proceed seems therefore as follows: 
From the theoretical side, and especially from lattice calculations
(see e.g. \cite{LAT1}), much is known about thermodynamical properties
of the QGP in a static situation. Although there is no direct possibility
to compare to experiment, this knowledge can be used under certain circumstances
as an input for dynamical fireball models, which constitute the next
important building block for modelling any specific process
inside such a system.
Here, experimental input is available in the form of bulk
hadronic observables such as directed flow and total multiplicities which characterize
the evolution endpoint. The third and final step is then the calculation
of a given process within the arena prepared by the fireball evolution.
In this way, fireball evolution and the process in question are tied to
different observables and are readily disentangled.

In the following, we try to elaborate on the second building
block, the dynamical fireball evolution in a simple framework
and calculate as an application dilepton radiation from this
fireball. This is more completely worked out in \cite{RENK}.

\section{Model framework}

We do not aim at the detailed description of the
heavy-ion collision or its subsequent expansion on
an event by event basis or by a hydrodynamical simulation.
Instead we use a model of an expanding fireball which
enables us to rapidly test different scenarios with
different parameter sets in a systematic way, so as
to gain insight into
the time evolution of the strongly interacting system.

We assume that the physics of the fireball is the same
inside each volume of given proper time $\tau$, thus
averaging over spatial inhomogenities in density and
temperature. The volume itself is taken to
be an expanding cylinder, in which the volume elements
move away from the center in order to generate the
observed flow. There is no global Lorentz frame in which
thermodynamics can be applied. As the fireball
expands, volume elements away from the center are
moving with large velocities and are subject to time
dilatation when seen in the center of mass frame
of the collision. 
In this frame, the fireball expands, at a given time,
much more rapidly
in the center than at the edges and does not resemble a cylinder
any more.
We assume a
linear increase in rapidity when going from central
volume elements to the fireball edge along the beam ($z$)-axis and the transverse axis. As the velocities along the
$z$-axis are typically large (up to $c$) as compared to transverse motion
(up to 0.55 $c$) for SPS and RHIC conditions, we make the
simplifying assumption that the proper time is in a
one-to-one correspondence to the $z$-position of a
given volume element, thus neglecting the time dilatation
caused by transverse motion. 
The whole system is assumed to
be in local thermal (though not necessarily chemical)
equilibrium at all times.

Given this overall framework, the volume expansion of the fireball
is governed by the longitudinal growth speed $v_z$ and the
transverse expansion speed  $v_\perp$ at a given proper time.
These quantities can be determined at the freeze-out point and
correspond to the observed amount of flow. However, flow
is measured in the lab frame and needs to be translated into
the growth of proper time volume.

Note that it is important to keep track of the velocity $v_z$ at
the fireball edge in this setup. If this quantity is simply
assumed to be $c$, the volume of given proper time does
not grow like $\tau c$ as one might naively expect (the c.m.
frame volume does so, however), but is infinite right from the
beginning.

We use a detailed analysis of the freeze-out conditions for
central Pb-Pb collisions at 160 AGeV \cite{FREEZE-OUT} to fix the endpoint of the
evolution. The initial state is constrained using the overlap
geometry of the colliding nuclei. The expansion between 
initial and freeze-out stages is
then required to be in accordance with the EoS as determined
from the quasiparticle model described in \cite{RAS}.

The volume expansion is parametrized by the following set of equations:
\begin{equation}
v_\perp(\tau) = \int_0^\tau d \tau' c_\perp\frac{p(\tau')}{\epsilon(\tau')} \qquad
R(\tau) = R_0 + \int_0^\tau \int_0^{\tau'} d \tau' d\tau'' c_\perp\frac{p(\tau'')}{\epsilon(\tau'')}
\end{equation}
\begin{equation}
v_z(t) = v_z^i + \int_0^t d t' c_z\frac{p(t')}{\epsilon(t')} \qquad
z(t)= z_0 + v_z^i \cdot t + \int_0^t \int_0^{t'} d t' dt'' c_z\frac{p(t'')}{\epsilon(t'')}.
\end{equation}
Here, the acceleration was assumed to be proportional to the ratio of pressure $p$
over energy density $\epsilon$ with a proportionality constant $c$.
The free parameters $c_\perp, c_z$, freeze-out proper time $\tau_f$ and 
freeze-out c.m. time $t_f$ can be fitted by requiring
agreement with initial conditions $R_0 \approx 4.5$ fm and $v_\perp$ = 0 
(overlap geometry) and final conditions
$R_f \approx 8.55$ fm, $\overline{v}_\perp = 0.5 c$,
$T_f = 100$ MeV, $v_z = 0.9 c$ (results from \cite{FREEZE-OUT}).
Assuming entropy conservation with an entropy per baryon of 26 for SPS
conditions at 160 AGeV, the entropy density $s$ at a given proper time
can then be obtained by dividing the total entropy $S_0$ by the volume,
$s = S_0/V(\tau)$. With the help of the EoS, the temperature $T(s)$, pressure
$p(s)$ and energy density $\epsilon(s)$ can then be calculated.
These are inserted into eqs.(1--2) in order to yield a self-consistent solution.
Chemical potentials for hadrons are introduced in order to agree with the
experimentally observed particle abundancies. For the above conditions,
a rise in the pion chemical potential $\mu_\pi$ up to 123 MeV towards
freeze-out is necessary.

\section{Results}

\begin{figure}[htb]
\epsfig{file=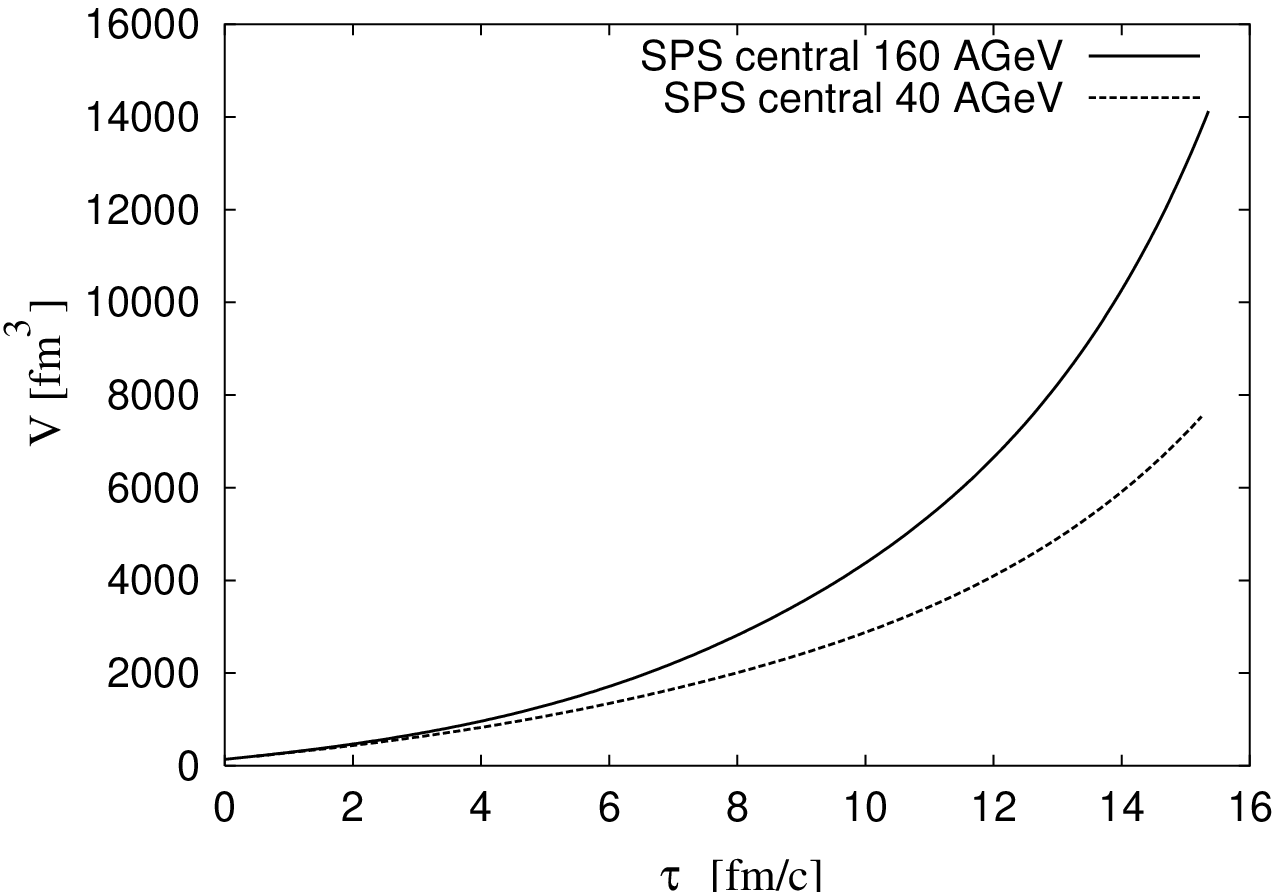, width=6.9cm}
\epsfig{file=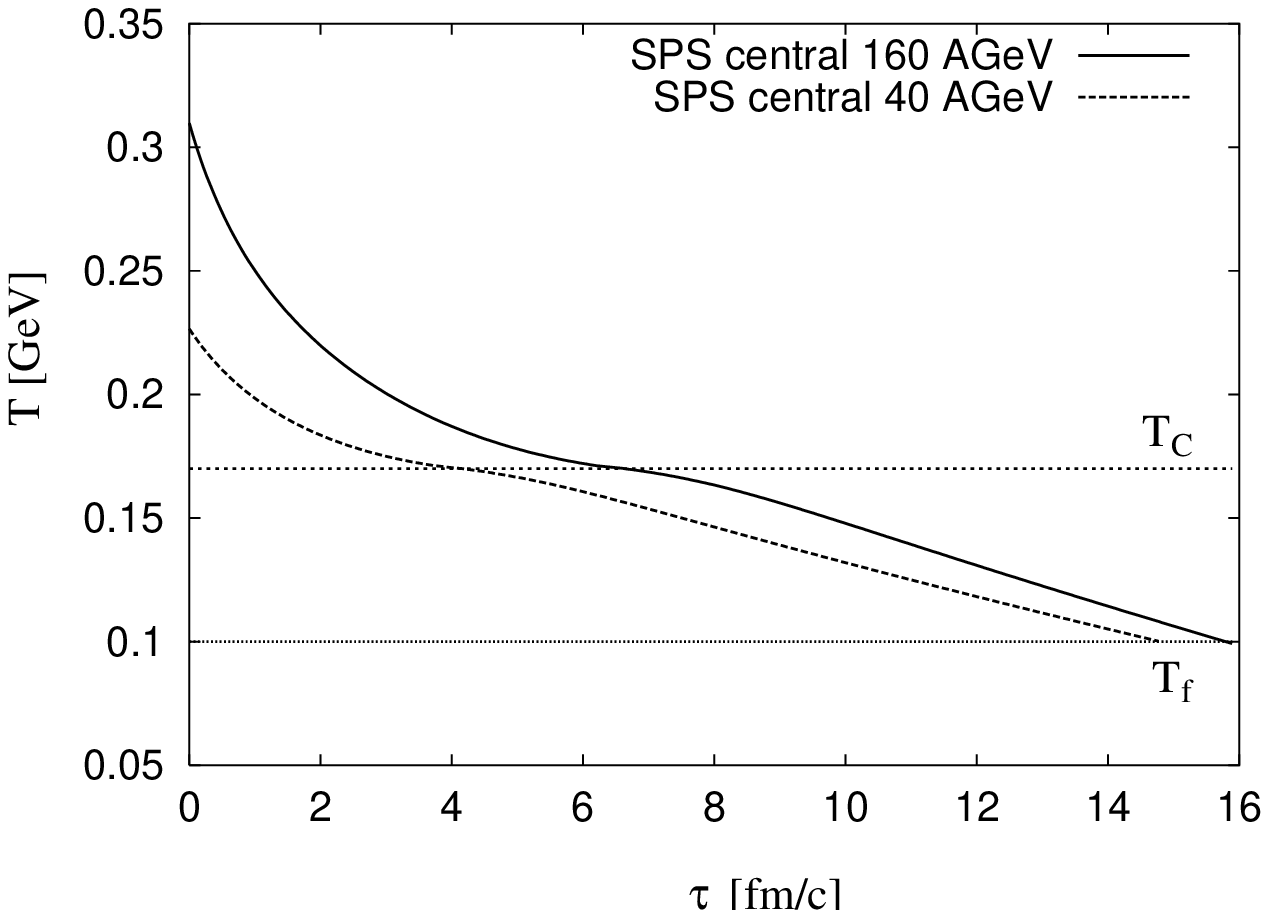, width=6.9cm}
\caption{\label{F-1}Left panel: Volume of constant proper time
at SPS 160 and 40 AGeV for central Pb-Pb collisions.
Right panel: Corresponding temperature profile with
entropy per baryon $s/\rho_B$ = 26 (160 AGeV) and 13 (40 AGeV).}
\end{figure}

The resulting volume expansion and temperature profile for
SPS conditions at two different energies is shown in
Fig.~\ref{F-1}.

The primary uncertainty is attached to the initial temperature, which
appears rather large ($\sim$300 MeV) for 160 AGeV collisions. 
Part of the difference in this value
as compared to other approaches comes from the use of a more
realistic equation of state, which even for the initial temparatures
differs considerably from the EoS of an ideal gas. Moreover, it
depends on the initial longitudinal expansion velocity.
After a few fm/c evolution time, however, the system is not
very sensitive to details of the initial conditions any more. 
Remarkably, even the 40 AGeV scenario shows some evolution above
the critical temperature $T_C = 170$ MeV. 
This is unavoidable --- for thermodynamical
reasons the volume required for an initial temperature below $T_C$ would
be by far too large to be found after the expansion within a  
sensible thermalization time of 1--2 fm/c.
Note that there is no mixed phase present. This is again a consequence
of the EoS which indicates a smooth crossover between
the two phases. Close to the critical temperature
there is a soft point in the EoS and its effect is
included in the model.

As a test for the extracted fireball evolution scenarios, we consider
dilepton emission. Once the thermal emission rate $\frac{dN(T(\tau),M, \eta,
p_T)}{d^4 x d^4p}$
from a hot source is known, the experimentally measured rate
can be calculated as

\begin{equation}
\frac{d^2N}{dM d\eta} =  \frac{2\pi M}{\Delta \eta} \int \limits_0^{\tau_{f}}
d\tau \  \int  d\eta \  V(T(\tau),\eta)
\int
\limits_0^\infty dp_T \ p_T
 \ \frac{dN(T(\tau),M, \eta,
p_T)}{d^4 x d^4p} \ Acc. \label{integratedrates}
\end{equation}

Here '$Acc$' refers to the acceptance characteristic of the detector.
Using the thermal quasiparticle model \cite{RAS} for the QGP spectral
function and and improved vector meson dominance model
combined with chiral dynamics \cite{KKW1} for the hadronic part,
we insert the resulting rate into the fireball evolution shown
above in order to compare to the CERES data \cite{CERES}.

\begin{figure}[htb]
\epsfig{file=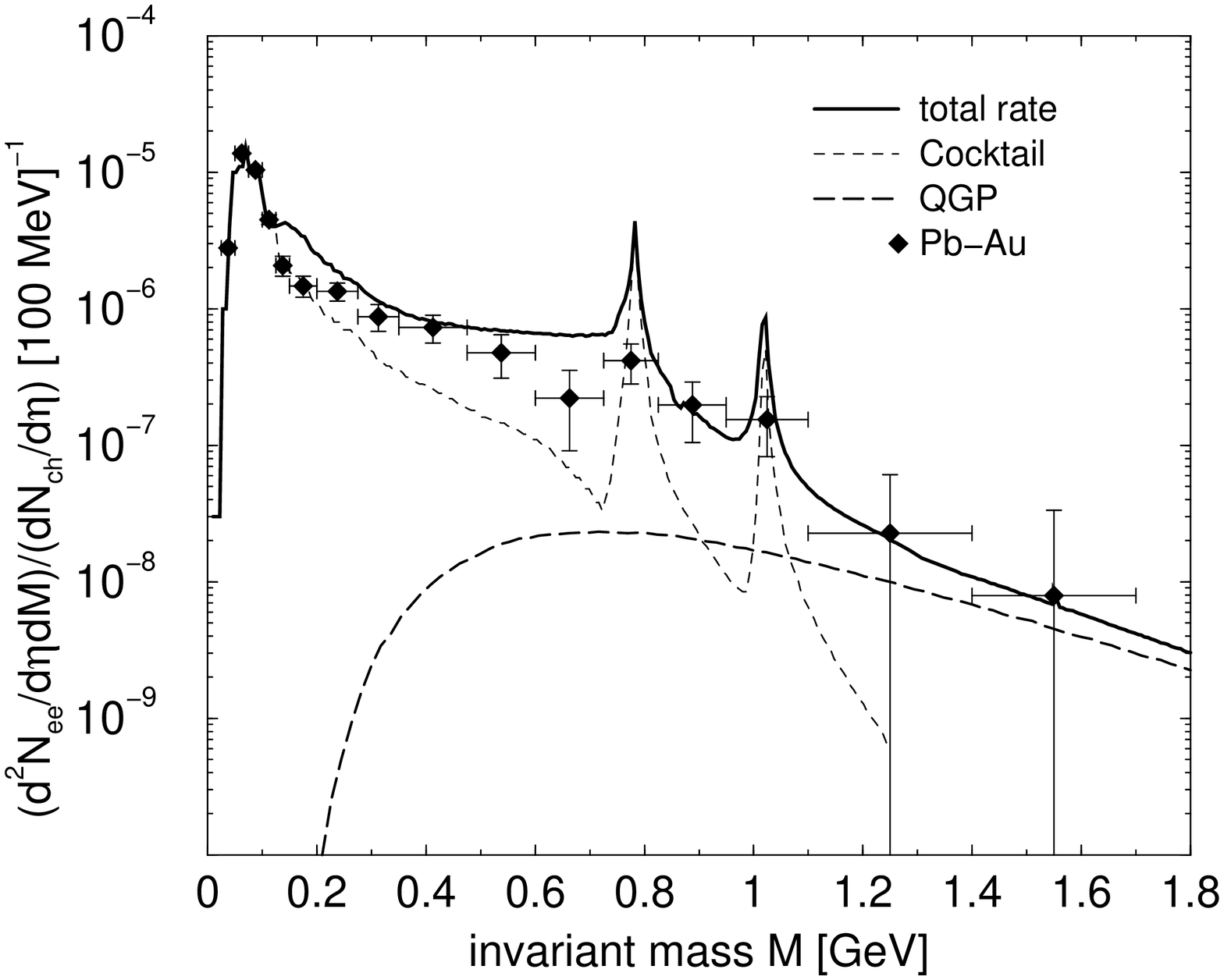, width=5.7cm, angle=-90} 
\epsfig{file=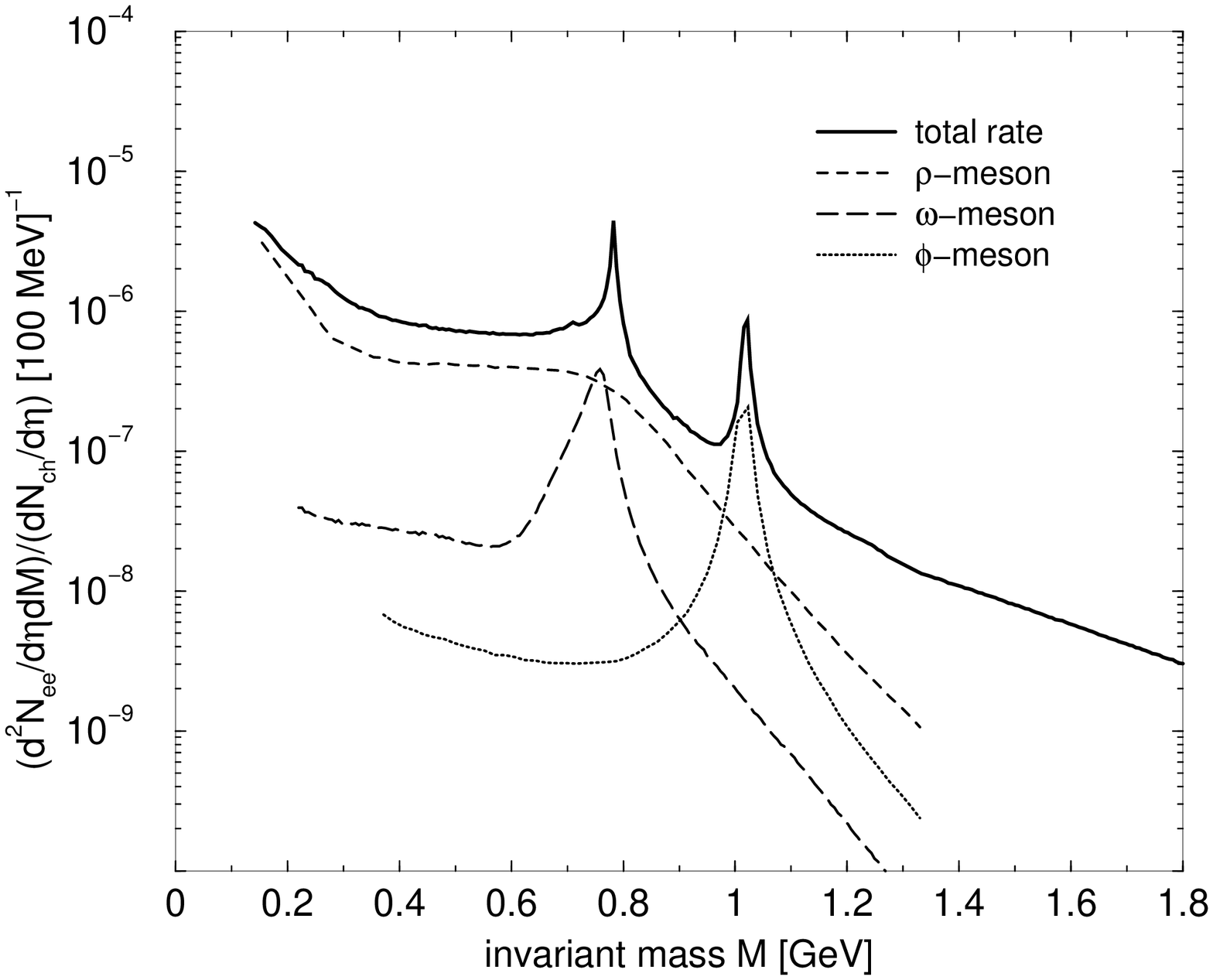, width=5.7cm, angle=-90}
\caption{\label{F-2}Left panel: Total dilepton yield for 30\% central
Pb-Au collisions at 160 AGeV (full), hadronic cocktail (dotted) and
QGP contribution (dashed). Right panel: Contribution of the vector
meson channels $\rho$, $\omega$ and $\phi$.}
\end{figure}

\begin{figure}[htb]
\epsfig{file=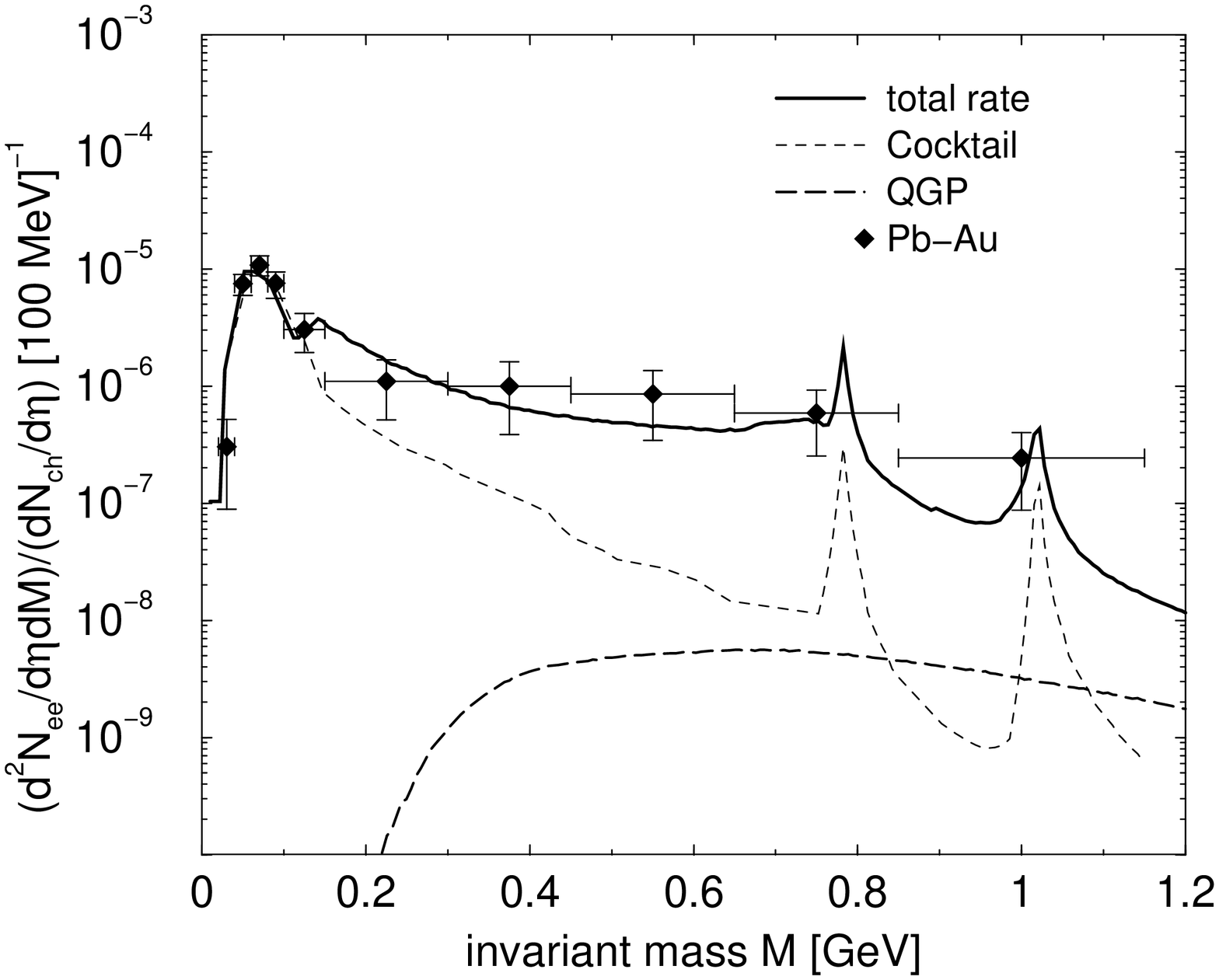, width=5.7cm, angle=-90} 
\epsfig{file=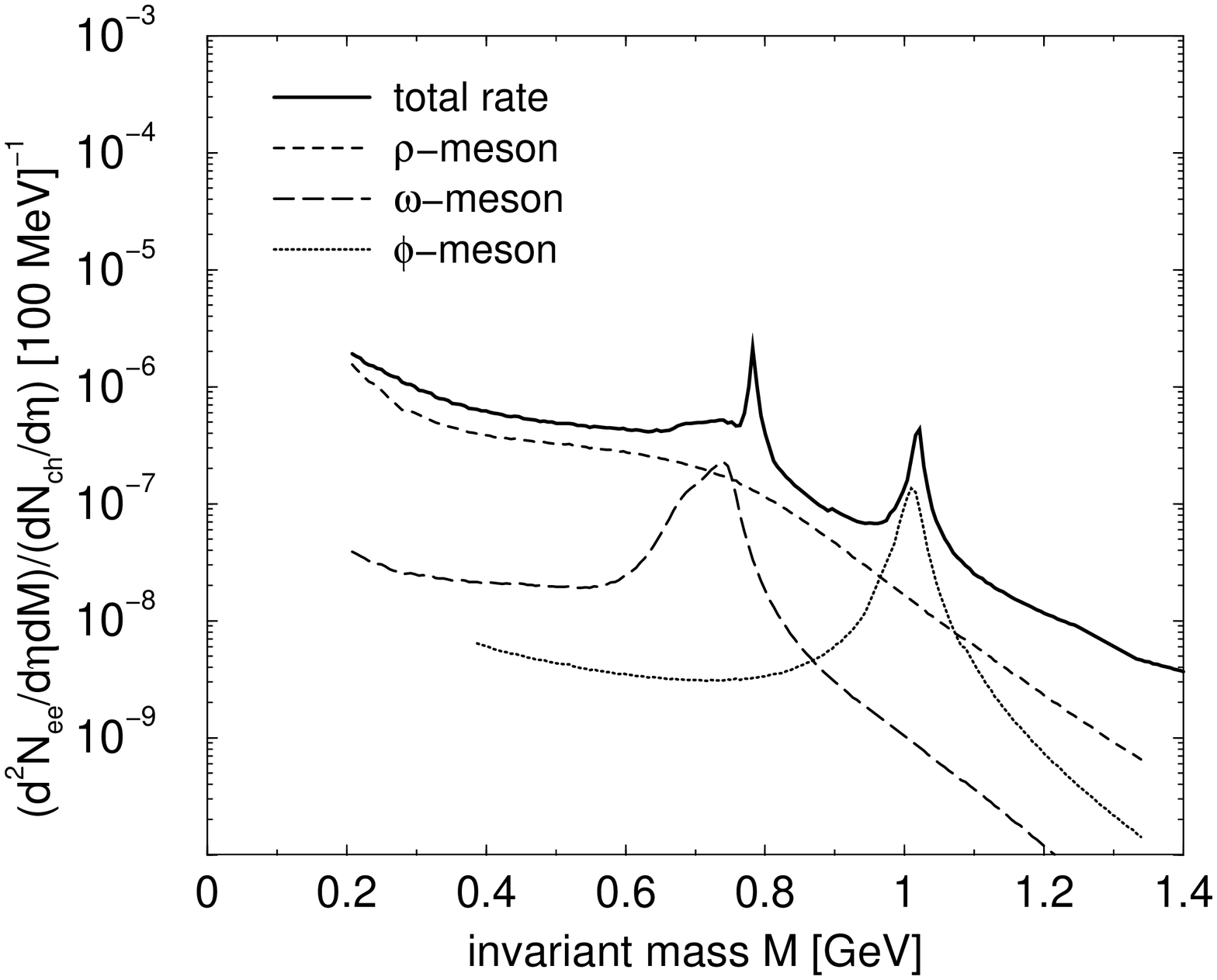, width=5.7cm, angle=-90}
\caption{\label{F-3}Left panel: Total dilepton yield for 30\% central
Pb-Au collisions at 40 AGeV (full), hadronic cocktail (dotted) and
QGP contribution (dashed). Right panel: Contribution of the vector
meson channels $\rho$, $\omega$ and $\phi$.}
\end{figure}

The 160 AGeV scenario is shown in Fig.~\ref{F-2}, the 40 AGeV scenario
in Fig.~\ref{F-3}. 
We calculate the full rate and also the contributions from different
processes, namely the thermal rate from the QGP, the so-called cocktail
contribution from Dalitz decays after freeze-out and the vector
meson channels. In both scenarios we find good agreement with the
data within errors. The driving force of the
dilepton excess around 500 MeV invariant mass is the broadening of
the $\rho$-meson due to finite baryon density.
The QGP contribution is visible at invariant masses above 1 GeV
in the 160 AGeV scenario and negligible for 40 AGeV.
Note that we do not aim for a best
fit. For example, in the invariant mass region around 200 MeV,
our calculation overshoots the data. This could be easily
cured by raising the freeze-out temperature which corresponds
to a reduction of the hadronic rate. However, as we choose
to fix the fireball evolution before with hadronic observables,
this is not an option any more.

It is clearly an interesting and challenging task to also apply the fireball
evolution model to other processes, e.g. direct photon emission
or J/$\Psi$ and see whether further agreement can be found, giving support
to the overall scenario. This is work currently in progress.


%
\end{document}